\documentstyle[aps,prl,rotate]{revtex}

\input{psfig}
\input{epsf}

\pssilent

\begin{document}

\title{ A First Measurement of Low $x$ Low $Q^{2}$ Structure Functions
in Neutrino Scattering}

\author{B.~T.~Fleming,$^{2}$ T.~Adams,$^{4}$ A.~Alton,$^{4}$
C.~G.~Arroyo,$^{2}$ S.~Avvakumov,$^{7}$ L.~de~Barbaro,$^{5}$
P.~de~Barbaro,$^{7}$ A.~O.~Bazarko,$^{2}$ R.~H.~Bernstein,$^{3}$
A.~Bodek,$^{7}$ T.~Bolton,$^{4}$ J.~Brau,$^{6}$ D.~Buchholz,$^{5}$
H.~Budd,$^{7}$ L.~Bugel,$^{3}$ J.~Conrad,$^{2}$ R.~B.~Drucker,$^{6}$
J.~A.~Formaggio,$^{2}$ R.~Frey,$^{6}$ J.~Goldman,$^{4}$
M.~Goncharov,$^{4}$ D.~A.~Harris,$ ^{7} $ R.~A.~Johnson,$^{1}$
J.~H.~Kim,$^{2}$ B.~J.~King,$^{2}$ T.~Kinnel,$^{8}$
S.~Koutsoliotas,$^{2}$ M.~J.~Lamm,$^{3}$ W.~Marsh,$^{3}$
D.~Mason,$^{6}$ K.~S.~McFarland, $^{7}$ C.~McNulty,$^{2}$
S.~R.~Mishra,$^{2}$ D.~Naples,$^{4}$ P.~Nienaber,$^{3}$
A.~Romosan,$^{2}$ W.~K.~Sakumoto,$^{7}$ H.~Schellman,$^{5}$
F.~J.~Sciulli,$^{2}$ W.~G.~Seligman,$^{2}$ M.~H.~Shaevitz,$^{2}$
W.~H.~Smith,$^{8}$ P.~Spentzouris, $^{2}$ E.~G.~Stern,$^{2}$
N.~Suwonjandee,$^{1}$ A.~Vaitaitis,$^{2}$ M.~Vakili,$^{1}$
U.~K.~Yang,$^{7}$ J.~Yu,$^{3}$ G.~P.~Zeller,$^{5}$ and
E.~D.~Zimmerman$^{2}$}

\address{( The CCFR/NuTeV Collaboration ) \\
$^{1}$ University of Cincinnati, Cincinnati, OH 45221 \\
$^{2}$ Columbia University, New York, NY 10027 \\
$^{3}$ Fermi National Accelerator Laboratory, Batavia, IL 60510 \\
$^{4}$ Kansas State University, Manhattan, KS 66506 \\
$^{5}$ Northwestern University, Evanston, IL 60208 \\
$^{6}$ University of Oregon, Eugene, OR 97403 \\
$^{7}$ University of Rochester, Rochester, NY 14627 \\
$^{8}$ University of Wisconsin, Madison, WI 53706\\ }

\maketitle

\begin{abstract}
A new structure function analysis of CCFR deep inelastic $\nu$-N and
$\overline{\nu}$-N scattering data is presented for previously
unexplored kinematic regions down to Bjorken $x=0.0045$ and
$Q^{2}=0.3$ GeV$^{2}$.  Comparisons to charged lepton scattering data
from NMC \cite{NMC} and E665\cite{E665} experiments are made and the
behavior of the structure function $F_{2}^{\nu}$ is studied in the
limit $Q^{2} \rightarrow 0$.
\vspace{0.1in} 
\end{abstract}


\newpage 
\narrowtext
\twocolumn

Neutrino structure function measurements in the low Bjorken $x$, low
$Q^2$ region can be used to study the axial-vector component of the
weak interaction as well as to test the limits of parton distribution
universality.  In this paper, we present a first measurement of the
structure function $F_2$ in neutrino scattering, from the CCFR data,
for $Q^2<1$ GeV$^{2}$ and $0.0045 < x < 0.035 $.  In this region where
perturbative and non-perturbative QCD meet, we present a
parameterization of the data which allows us to test the partially
conserved axial current (PCAC) limit of $F_2$ in neutrino scattering.

The universality of parton distributions can be tested by comparing
neutrino scattering data to charged lepton scattering data.  Past
measurements for $0.0075<x<0.1$ and $Q^{2}>1.0$ GeV$^{2}$ have
indicated that $F_{2}^{\nu }$ differs from $F_{2}^{\mu }$ by 10-15\%
\cite{seligman:1997mc}.  This discrepancy has been partially resolved
by recent analyses of $F_{2}^{\nu}$ at $Q^2 > 1.0 $
GeV$^{2}$ \cite{unki,thomas}.  While we expect and have now observed that
parton distribution universality holds in this region, this need not
be the case at lower values of $Q^{2}$.  Deviations from this
universality at lower $Q^{2}$ are expected due to differences in
vector and axial components of electromagnetic and weak interactions.
In particular, the electromagnetic interaction has only a vector
component while the weak interaction has both vector and axial-vector
components.  Vector currents are conserved (CVC) but axial-vector
currents are only partially conserved (PCAC).  Adler \cite{adler}
proposed a test of the PCAC hypothesis using high energy neutrino
interactions, a consequence of which is the prediction that $F_{2}$
approaches a non-zero constant as $Q^{2} \rightarrow 0$ due to U(1)
gauge invariance.  A determination of this constant is performed here
by fitting the low $Q^{2}$ data to a phenomenological curve developed
by Donnachie and Landshoff \cite{DL}.

The differential cross-sections for the $\nu$N charged-current process
$\nu _{\mu }\left( \overline{\nu}_{\mu }\right) +N\rightarrow \mu
^{-}\left( \mu ^{+}\right) +X$ in the limit of negligible quark masses
and neglecting lepton masses, in terms of the Lorentz-invariant
structure functions $F_{2} $, $2xF_{1}$, and $xF_{3}$ are:
\begin{eqnarray}
\frac{d\sigma ^{\nu ,\overline{\nu }}}{dx\ dy} &=& \frac{G_{F}^{2}ME_{\nu }}{
\pi }\left[ \left( 1-y-\frac{Mxy}{2E_{\nu }}\right) F_{2}\left(
x,Q^{2}\right) \right. \nonumber \\
&&\left. +\frac{y^{2}}{2}2xF_{1}\left( x,Q^{2}\right) \pm y\left( 1-%
\frac{y}{2}\right) xF_{3}\left( x,Q^{2}\right) \right]  \label{eq:dxdy}
\end{eqnarray}
where $G_{F}$ is the weak Fermi coupling constant, $M$ is the nucleon
mass, $E_{\nu }$ is the incident $\nu$ energy, $Q^{2}$ is the square
of the four-momentum transfer to the nucleon, the scaling variable
$y=E_{HAD}/E_{\nu }$ is the fractional energy transferred to the
hadronic vertex with $E_{HAD}$ equal to the measured hadronic energy,
and $x=Q^{2}/2ME_{\nu }y$, the Bjorken scaling variable, is the
fractional momentum carried by the struck quark. The structure
function $2xF_{1}$ is expressed in terms of $F_{2}$ by
$2xF_{1}(x,Q^{2})=F_{2}(x,Q^{2})\times
\frac{1+4M^{2}x^{2}/Q^{2}}{1+R(x,Q^{2})}$, where $R=\frac{\sigma
_{L}}{\sigma _{T}} $ is the ratio of the cross-sections of
longitudinally to transversely-polarized $W$ bosons. In the leading
order (LO) quark-parton model, $F_{2}$ is the sum of the momentum
densities of all interacting quark constituents, and $xF_{3}$ is the
difference of these, the valence quark momentum density; these
relations are modified by higher-order QCD corrections.

The $\nu$ deep inelastic scattering (DIS) data were collected in two
high-energy high-statistics runs, FNAL E744 and E770, in the Fermilab
Tevatron fixed-target quadrupole triplet beam (QTB) line by the CCFR
collaboration. The detector~\cite{hadcal,mucal} consists of a target
calorimeter instrumented with both scintillators and drift chambers
for measuring the energy of the hadron shower, $E_{HAD}$, and the muon
angle, $\theta _{\mu }$, followed by a toroid spectrometer for
measuring the muon momentum $p_{\mu }$. There are 1,030,000 $\nu _{\mu
}$ events and 179,000 $\overline{\nu }_{\mu }$ events in the data
sample after fiducial volume and geometric cuts, and kinematic cuts
of $p_{\mu }$ $>15\ $GeV, $\theta _{\mu }<150\ $mr, $E_{HAD}$ $>10\
$GeV, and $30<E_{\nu }<360\ $GeV.  These cuts were applied to
select regions of high efficiency and small systematic errors in
reconstruction.

The structure function $F_{2}$ in Eq.~(\ref{eq:dxdy}) can be
calculated from the observed number of $\nu _{\mu }$ and
$\overline{\nu }_{\mu }$ events combined with the $\nu _{\mu }$ and
$\overline{\nu }_{\mu }$ fluxes.  The ratio of
fluxes between different energies in $\nu$ mode and that between
$\overline{\nu}$ and $\nu$ mode was determined using the events with
$E_{HAD} < 20$GeV ~\cite{WGSthesis,Auc,BelRein}.  The overall
normalization of the flux was constrained such that the measured total
neutrino-nucleon cross-section for $\nu$s equaled the world average
cross-section for isoscalar-corrected iron target experiments, $\sigma
^{\nu \,Fe}/E=(0.677\pm 0.014)\times 10^{-38}$cm$^{2}/$GeV
\cite{Auc,worldsig} and for $\overline{\nu}$s equaled the world
average cross-section including this experiment for
isoscalar-corrected iron target experiments, $\sigma ^{\overline{\nu}
\,Fe}/E=(0.340\pm 0.007)\times 10^{-38}$cm$^{2}/$GeV.  Negligible
corrections for non-isoscalarity of the iron target and the mass of
the $W$ boson propagator are applied.

Sources of systematic error on $F_{2}$ arise from limitations of the
models used for corrections and from the level of our knowledge of the
detector calibration.  Muon and hadron energy calibrations were
determined from test beam data collected during the course of the
experiment \cite{hadcal,mucal}. For acceptance, smearing, and
radiative corrections we chose an appropriate model for the low $x$,
low $Q^{2}$ region, the GRV \cite{GRV} model of the parton
distribution functions.  The GRV model is used up to $Q^{2}=1.35$
GeV$^{2}$ where it is normalized to a LO parameterization \cite{BG}
used above this.  Inclusion of the GRV model in the radiative
correction calculation caused a systematic decrease in $F_{2}$ by as
much as 10\% in the lowest $x$ bin, decreasing to 1-2\% at $x=0.015$
as compared to the effects of the LO model used in the previous
analysis \cite{WGSthesis}.  Due to the systematic uncertainty in the
model at low $x$, the radiative correction error is 3\% in the lowest
$x$ bin.  A correction is applied for the difference between
$xF_{3}^{\nu}$ and $xF_{3}^{\overline{\nu}}$, determined using a LO
calculation of $\Delta xF_{3}= xF_{3}^{\nu} - xF_{3}^ {
\overline{\nu}}$.  The recent CCFR $\Delta xF_{3}$ measurement
\cite{unki} is higher than this LO model \cite{BG} and all other
current LO and NLO theoretical predictions in this kinematic region.
An appropriate systematic error is applied to account for the
differences between the theory and this measurement. Finally, a
systematic error is applied to account for the uncertainty in the
value of $R$, which comes from a global fit to the world's
measurements \cite{Rworld}.

In previous analyses a slow rescaling correction was applied to
account for massive charm effects.  This is not applied here since the
corrections are model dependent and uncertain in this kinematic range.
As a result, neutrino and charged lepton DIS data must be compared
within the framework of charm production models, accomplished by
plotting the ratio of data to theoretical model.  The theoretical
calculation corresponding to the CCFR data employs NLO QCD including
heavy flavor effects as implemented in the TR-VFS(MRST99) scheme
\cite{mfs,mrst}.  The theoretical calculation corresponding to NMC and
E665 data is determined using TR-VFS(MRST99) for charged lepton
scattering.  Other theoretical predictions such as ACOT-VFS(CTEQ4HQ)
\cite{vfs,acot} and FFS(GRV94) \cite{ffs} do not significantly change
the comparison.

\begin{figure}[tbp]
\centerline{\psfig{figure=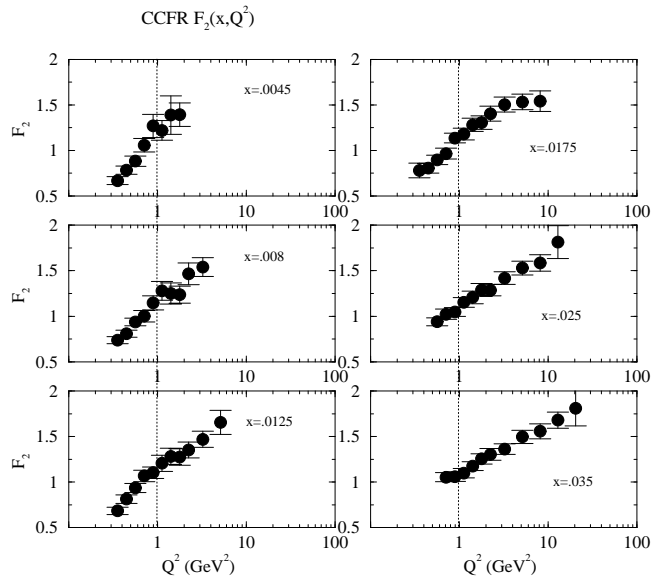,width=3.375in}}
\caption{CCFR $F_{2}$ at low $x$, low $Q^{2}$.  Data to the left of the vertical line at $Q^{2}=1.0$ represent the new kinematic regime for this analysis.}
\label{fig:SF}
\end{figure}

\begin{figure}[tbp]
\centerline{\psfig{figure=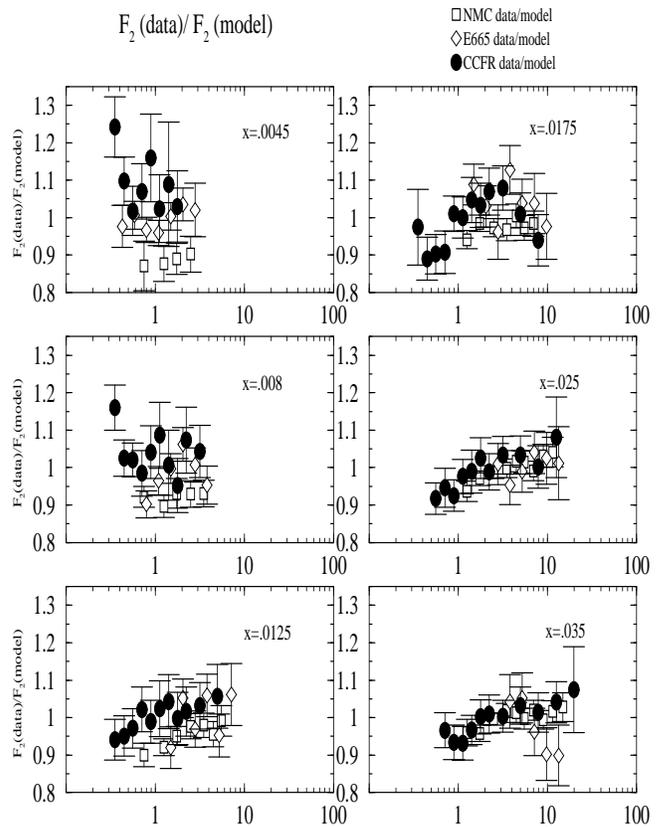,width=3.375in,height=4.3in}}
\caption{$F_{2}$ data/theory from CCFR $\nu $-Fe DIS compared to
$F_{2}$ from NMC and E665 DIS. Errors bars are statistical and
systematic added in quadrature. Theoretical predictions are those of
TR-VFS(MRST99).}
\label{fig:compare}
\end{figure}

The combination of the inclusion of the GRV model at low $x$ and low
$Q^{2}$, its effect on the radiative corrections, and removal of the
slow rescaling correction help to resolve the longstanding discrepancy
between the neutrino and charged lepton DIS data
above $x=0.015$. $F_{2}$ is plotted in Figure~\ref{fig:SF}.  Errors
are statistical and systematic added in quadrature. A line is drawn at
$Q^{2}=1$ GeV$^{2}$ to highlight the kinematic region this analysis
accesses.  Figure \ref{fig:compare} compares $F_{2}$ (data/theoretical
model) for CCFR, NMC, and E665.  There is agreement to within 5\% down
to $x=0.0125$.  Below this, as $x$ decreases, CCFR $F_{2}$
(data/theory) becomes systematically higher than NMC $F_{2}$
(data/theory). Differences between scattering via the weak interaction
and via the electromagnetic interaction as $Q^{2} \rightarrow 0$ may
account for the disagreement in this region.

\begin{figure}[tbp]
\centerline{\psfig{figure=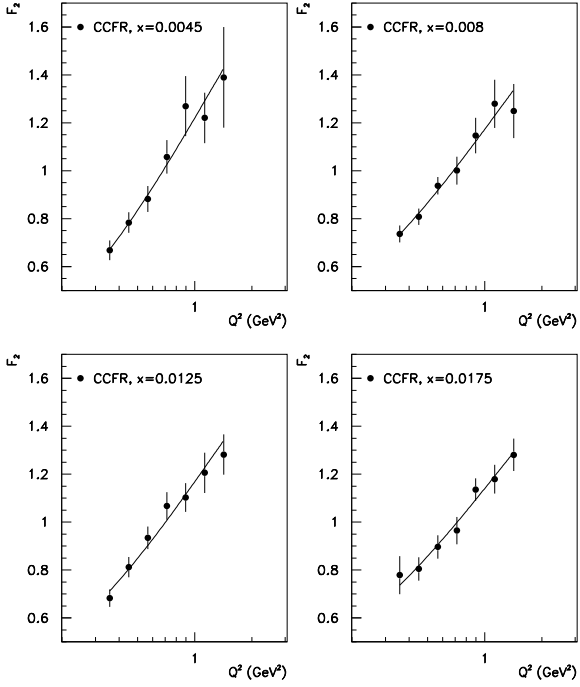,width=3.375in}}
\caption{Results from fit to CCFR data to extrapolate to $F_{2}(Q^{2}=0)$.}
\label{fig:fits}
\end{figure}

\begin{figure}[tbp]
\centerline{\psfig{figure=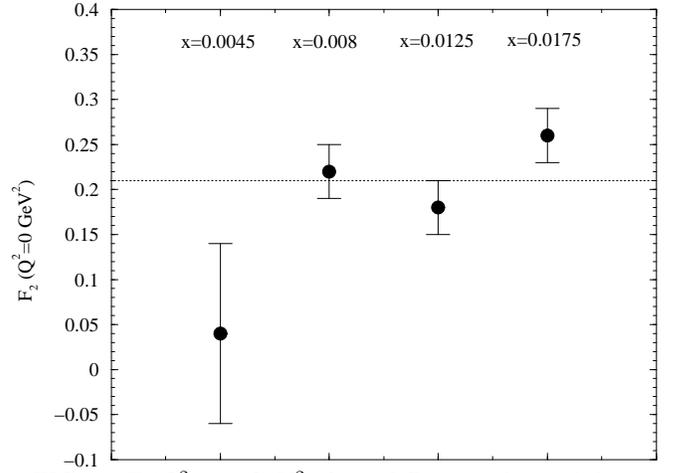,width=3.375in}}
\caption{$F_{2}(Q^{2}=0$ GeV$^{2})$ from different $x$ bins.  A line is drawn at the weighted average of all four measurements.}
\label{fig:fitsplot}
\end{figure}

In charged lepton DIS, the structure function
$F_{2}$ is constrained by gauge invariance to vanish with
$Q^{2}$ as $Q^{2}\rightarrow 0$.  Donnachie and Landshoff predict that in the
low $Q^{2}$ region, $F_{2}^{\mu}$ will follow the form \cite{DL}:
\begin{equation}
C \left( \frac{Q^{2}}{Q^{2}+A^{2}} \right) .
\label{eq:nmc_e665}
\end{equation}
However, in the case of neutrino DIS, the axial
component of the weak interaction may contribute a nonzero component to
$F_{2}$ as $Q^{2}$ approaches zero.  Donnachie and Landshoff predict
that $F_{2}^{\nu}$ should follow a form with a non-zero contribution
at $Q^{2}=0$:
\begin{equation}
\frac{C}{2} \left( \frac{Q^{2}}{Q^{2}+A^{2}} + \frac{Q^{2} + D}{Q^{2}
+ B^{2}} \right) .
\end{equation}  
Using NMC and E665 data, corrected in this case to be equivalent to
scattering from an iron target using a parameterization of SLAC Fe/D
data \cite{WGSthesis}, we do a combined fit to the form predicted for
$\mu$ DIS and extract the parameter $A = 0.81
\pm 0.02$ with $\chi ^{2}/DOF = 27/17$.  Results of fits in each
$x$ bin for each experiment are shown in Table \ref{tab:A_fitresults}
for comparison to parameters in the CCFR fit.  The error on $A$ is
incorporated in the systematic error on the final fit.  Inserting this
value for $A$ into the form predicted for $\nu$N DIS, we fit CCFR data to extract parameters B, C, and D, and
determine the value of $F_{2}$ at $Q^{2}=0$.  Only data below
$Q^{2}=1.4$ GeV$^{2}$ are used in the fits.  The CCFR $x$-bins that
contain enough data to produce a good fit in this $Q^{2}$ region are
$x=0.0045$, $x=0.0080$, $x=0.0125$, and $x=0.0175$.  Figure
\ref{fig:fits} and Table \ref{tab:fitresults} show the results of the
fits.  Error bars consist of statistical and systematic terms added in
quadrature but exclude an overall correlated normalization uncertainty
of 1-2\%. The values of $F_{2}$ at $Q^{2}=0$ GeV$^{2}$ in the three
highest $x$-bins are statistically significant and are within $1
\sigma$ of each other.  The lowest $x$ bin has large error bars but is
within $1.5 \sigma$ of the others.  Taking a weighted average of the
parameters $B, C, D,$ and $F_{2}$ yields $B=1.53 \pm 0.02, C=2.31 \pm
0.03, D=0.48 \pm 0.03$, and $F_{2}(Q^{2}=0)=0.21 \pm 0.02$.  Figure
\ref{fig:fitsplot} shows $F_{2}(Q^{2}=0)$ for the different $x$ bins.
Inclusion of an $x$ dependence of the form $x^{\beta}$ does not change
the overall fits or $\chi ^{2}$s.  However, unlike in charged lepton
scattering, the Donnachie and Landshoff mass parameter, $B$, appears
to depend on $x$, with higher values corresponding to higher $x$.
Thus, $F_{2}$ at higher $x$ approaches $F_{2}(Q^{2}=0)$ more slowly
than at lower $x$.

In summary, a comparison of $F_{2}$ from neutrino DIS to that from
charged lepton DIS shows good agreement above $x=0.0125$, but shows
differences at smaller $x$. This low $x$ discrepancy can be explained
by the different behavior of $F_{2}$ from $\nu$ DIS to that from
$e/\mu$ DIS as $Q^{2} \rightarrow 0$.  CCFR $F_{2}^{\nu}$ data favors
a non-zero value for $F_{2}$ as $Q^{2} \rightarrow 0$.

We would like to thank the management and staff of Fermilab, and
acknowledge the help of many individuals at our home institutions.  We
would also like to thank Fred Olness for many useful discussions \cite{fred}.
This research was supported by the National Science Foundation and the
Department of Energy of the United States, who should be credited for
their continued support of basic research.

\begin{table}[tbp]
\caption{Results for NMC and E665 data fit to Eq.~\ref{eq:nmc_e665}.}
\begin{minipage}[t]{0.5\textwidth}
\begin{tabular}{cccc}
$x$ & $A$ & $\chi^{2}/N$ & $N$\\
\hline \hline $0.0045 (NMC)$ & $0.87 \pm 0.16$ & $0.02$ & $2$
\\$0.0045 (E665)$\footnote{Bin center corrected from $x=0.004$} & $0.90 \pm 0.10$ & $0.43$ & $4$
\\$0.0045 (E665)$\footnote{Bin center corrected from $x=0.005$} & $0.94 \pm 0.09$ & $0.31$ & $5$
\\$0.0080 (NMC)$ & $0.75 \pm 0.07$ & $0.38$ & $3$
\\$0.0080 (E665)$\footnote{Bin center corrected from $x=0.007$} & $0.87 \pm 0.10$ & $0.24$ & $4$
\\$0.0080 (E665)$\footnote{Bin center corrected from $x=0.009$} & $0.85 \pm 0.11$ & $1.19$ & $4$
\\$0.0125 (NMC)$ & $0.81 \pm 0.05$ & $0.55$ & $5$ 
\\$0.0125 (E665) $ & $0.97 \pm 0.14$ & $1.12$ & $4$ 
\\$0.0175 (NMC)$ & $0.78 \pm 0.06$ & $0.38$ & $5$
\\$0.0175 (E665) $ & $0.76 \pm 0.13$ & $0.88$ & $5$\\
\end{tabular} 
\end{minipage}
\label{tab:A_fitresults}
\end{table}

\begin{table}[t]
\caption{Fit results for CCFR data. CCFR data is fit to Eq. 4 with $A =
0.81 \pm 0.02$ as determined by fits to NMC and E665 data.  B, C, D, and
$F_{2}$ at $ Q^{2}=0$ results shown below. $N=4$ for all fits.}
\begin{center}
\begin{tabular}{ccccccc}
$x$ & $B$ & $C$ & $D$ & $F_{2}^{\nu}(Q^{2}=0)$ & $\chi^{2}/N$\\
\hline \hline $0.0045$ & $1.49 \pm 0.02$ & $2.62 \pm 0.26$ & $0.06
\pm 0.17$ & $0.04 \pm 0.10$ & $0.5$\\ 
$0.0080$ & $1.63 \pm 0.05$ & $2.32 \pm 0.05$ & $0.50 \pm 0.05$ & $0.22 \pm 0.03$ & $0.5$\\
$0.0125$ & $1.63 \pm 0.05$ & $2.39 \pm 0.05$ & $0.40 \pm 0.05$ & $0.18
\pm 0.03$ & $1.0$\\ $0.0175$ & $1.67 \pm 0.05$ & $2.20 \pm 0.05$ &
$0.65 \pm 0.07$ & $0.26 \pm 0.03$ & $0.5$\\
\end{tabular} 
\end{center}
\label{tab:fitresults}
\end{table}

\end{document}